\documentclass[11pt]{article}
\usepackage{amsmath,amsfonts,sint,epsfig,cite,graphics}
\usepackage{mathrsfs}
\usepackage{macros_static}
\usepackage{wrapfig}
\usepackage{color}

\usepackage[english]{babel}

\usepackage{psfrag}
\usepackage{amsmath}
\usepackage{dsfont}
\usepackage{subfigure}
\usepackage{axodraw}
\usepackage{feynmp}
\usepackage{amsfonts}

\newcommand{\bdm}{\begin{displaymath}}
\newcommand{\edm}{\end{displaymath}}
\renewcommand{\be}{\begin{equation}}
\renewcommand{\ee}{\end{equation}}
\renewcommand{\bi}{\begin{itemize}}
\renewcommand{\ei}{\end{itemize}}

\renewcommand{\SUthree}{{SU(3)}}

\renewcommand{\SUtwo}{{SU(2)}}

\renewcommand{\Tr}{{\rm Tr}}
\newcommand{\Str}{{\rm Str}}

\include{include}

\begin{document}

\renewcommand{\vec}[1]{{\bf #1}}
\newcommand{\PH}{P_{\rm H}}
\newcommand{\PL}{P_{\rm L}}
\newcommand{\unity}{{I}}

\def\slash#1{\mkern-1.5mu\raise0.6pt\hbox{$\not$}\mkern1.2mu #1\mkern 0.7mu}
\begin{titlepage}

\begin{flushright}
\vskip .4cm
HIM-2010-01\\
MKPH-T-10-17\\
CERN-PH-TH/2010-205\\
\end{flushright}

\vskip 2.25cm
\begin{center}
{\Large\bf 
Quark disconnected diagrams in \\[1mm] chiral perturbation theory}
\end{center}
\vskip 1.1cm
\begin{center}
{
Michele~Della~Morte$^{\scriptscriptstyle a}$ and
Andreas J\"uttner$^{\scriptscriptstyle b}$

\vskip 1cm
}
\vskip 1.3cm
{
\vskip 2.0ex
$^{\scriptstyle a}$
Institut~f{\"u}r~Kernphysik and Helmholtz Institut Mainz, \\
Johannes Gutenberg-Universit{\"a}t, D-55099 Mainz, 
Germany
\vskip 2.0ex
$^{\scriptstyle b}$
CERN, Physics Department, TH Unit, CH-1211 Geneva 23, Switzerland
}
\end{center}
\vskip 1.3cm
{\bf Abstract:}
We show how
quark-disconnected and quark-connected contributions  
to hadronic $n$-point functions can be written as independent correlators
for which one can derive expressions in partially
quenched chiral effective theory.
As an example we apply the idea to the case of the hadronic vacuum polarisation.
In particular, we consider
the cases of the $N_f=2$ theory without and with a partially quenched
strange quark and also the $N_f=2+1$ theory. In the latter two cases a 
parameter-free prediction for the disconnected contribution at NLO in the
effective theory is given. 
Finally we show how twisted boundary conditions
 can then be used in lattice QCD to improve the $q^2$ resolution in the
 connected contributions even when flavour singlet operators are
 considered.

\end{titlepage}

\newpage
\tableofcontents
\section{Introduction}
Hadronic correlation functions which involve quark disconnected
contributions are notoriously hard to compute in lattice 
QCD~\cite{Neff:2001zr,Foley:2005ac}.
Such contributions however appear in several 
 quantities related to properties of flavour singlet particles 
(e.g. the $\eta$ and $\eta'$ mesons) or to matrix elements of 
flavour singlet operators (as for the strangeness content of the nucleon),
 or to electro-magnetic interactions (e.g. the hadronic contribution to
 the anomalous magnetic moment of the muon or to the nucleon electric 
dipole moment).
There is no conceptual difficulty in treating disconnected quark diagrams
but the computational effort is immense
compared to quark-connected contributions. The disconnected part is 
therefore often neglected, not always providing solid arguments that the
systematic uncertainty introduced in this way is under control.

Here we present a method that allows to predict the quark disconnected
contribution to correlation functions in (partially quenched)
chiral perturbation theory \cite{Gasser:1983yg,Gasser:1984gg,
Bernard:1992mk,Bernard:1993sv,Sharpe:2000bc}.
We first give the general argument and then present an explicit example
by deriving predictions for the magnitude of the quark disconnected 
contribution to the hadronic vacuum polarisation of the photon for
$N_f=2$ dynamical flavours, for $N_f=2$ dynamical flavours 
with a quenched strange quark and also for 
the case of $N_f=2+1$ active flavours.
The vacuum polarisation is the main ingredient in
computations of the leading hadronic contribution to the anomalous
magnetic moment of the muon and hence of relevance to precision tests of
the Standard Model (see Ref.~\cite{Jegerlehner:2009ry} for a review of the 
subject).

We will stay in the Euclidean continuum and infinite volume
for most of the discussion. While having applications to lattice QCD in mind,
reference to this regularisation is only made where we think that
it helps the better understanding of our arguments. We will
discuss however the finite volume case in order to
show how this technique allows for using  
{\it partial twisting} \cite{Sachrajda:2004mi,Bedaque:2004ax,Bedaque:2004kc,deDivitiis:2004kq,Tiburzi:2005hg,Flynn:2005in} also for form-factors and polarisations 
involving flavour-diagonal operators.

\section{General argument}

We consider a $n$-point fermionic correlation function in QCD with 
$N_f$ 
dynamical flavours (not necessarily degenerate). In general several 
Wick contractions contribute 
to the  correlator. By  introducing valence quarks which are degenerate 
with the dynamical flavours, each Wick contraction can be rewritten in terms
of a single fermionic correlation function defined in an un-physical 
theory. The 
physical result is recovered by summing over the correlation functions 
in the un-physical, partially quenched, theory 
\cite{Bernard:1992mk,Bernard:1993sv,Sharpe:2000bc}. 
In particular, the 
approach can be  used to separate the contributions from quark 
disconnected diagrams from those coming from quark connected diagrams.
These contributions taken on their own are un-physical, it is therefore
natural that in order to define them as correlation functions one has
to resort to un-physical theories.
Investigations of
quark-disconnected
diagrams along these lines can also be found in
Ref.~\cite{Sharpe:2000bc}.

The number of valence quarks $N_v(i)$ degenerate with the $i$th
dynamical flavour, which has to be introduced depends on the particular 
correlation function. Given a $n$-point fermionic correlation function,
$N_v(i)$ is related in an obvious way to the largest number
$N_{\rm D}^{\rm max}(i)$ of disconnected quark loops involving
the $i$th flavour, which can appear as all possible Wick contractions 
are considered. In particular $N_v(i)=N_{\rm D}^{\rm max}(i)-1$ at most,
in such a way that a different {\it flavour} can be {\it
attached} to each disconnected quark loop.
In this approach  the partially quenched theory, in which 
each Wick contraction of the original correlator can be 
written as an  independent correlation function, 
is not unique but rather depends on the specific $n$-point 
function we started from. 

Partially quenched chiral perturbation theory (PQ$\chi$PT) 
\cite{Bernard:1992mk,Bernard:1993sv,Sharpe:2000bc}, which is an extension
of chiral perturbation theory \cite{Gasser:1984gg,Gasser:1983yg},
provides 
an asymptotic low energy description of partially quenched QCD (PQQCD) 
and can therefore be used to obtain
predictions for, and algebraic relations amongst,
the disconnected and the connected part of a 
fermionic correlation function. There is a rather large body of literature on
PQ$\chi$PT and PQQCD, we found the reviews in Ref.~\cite{Sharpe:2006pu}
 and Ref.~\cite{Golterman:2009kw} very clear and useful.
In short, following Ref.~\cite{Sharpe:2006pu}, PQQCD can be formulated 
in  terms of a local theory by introducing a commuting spin-1/2 field,
a ghost \cite{Morel:1987xk}, labelled by $\tilde{q}$ for each valence quark $q$ and by
extending the fermionic QCD action by including terms of the form
$\bar{q}(\slash{{\rm D}}+m_q)q+\tilde{q}^{\dagger} 
(\slash{{\rm D}}+m_q)\tilde{q}$, which violate the spin statistics theorem. As long as all masses are positive, the
determinant produced by the integral over the valence quark fields is
cancelled by the corresponding integral over the ghost fields and the
QCD partition function is reproduced. The advantage is that  this 
formulation provides field-theoretic expressions for partially quenched
correlation functions. As discussed in 
Refs.~\cite{Sharpe:2006pu,Golterman:2009kw}, 
a low energy description of the theory (PQ$\chi$PT) is
obtained by constructing a chiral Lagrangian encoding the apparent
chiral symmetry group of the extended action, which is a
$SU\hspace{-.5mm}\left(N_f+\sum_{i=1}^{N_f}N_v(i)\;|\;  
\sum_{i=1}^{N_f}N_v(i)\right)$
graded group.

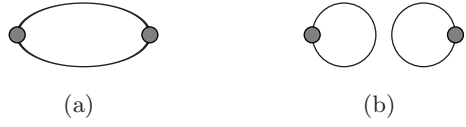
\begin{figure}
	\centering
	\subfigure[]{
	 \begin{picture}(60,15)(-30,-15) 
	 \Oval(0,0)(12,25)(0)
	 \GCirc(-25,0){3}{0.5}\GCirc(25,0){3}{0.5} 
	\end{picture}
	}
	\hspace{15mm}\subfigure[]{
	 \begin{picture}(60,15)(-30,-15) 
	 \Oval(-15,0)(12,12)(0)
	 \Oval(15,0)(12,12)(0)
	 \GCirc(-27,0){3}{0.5}\GCirc(27,0){3}{0.5} 
	\end{picture}
	}
	\caption{Wick contractions: (a) connected and (b) disconnected
		diagram.}
	\label{fig:Wicks}
\end{figure}

For the application discussed here the strategy can be exemplified
for the case of a meson 2-point function
constructed of flavour-diagonal quark-bilinears,
\begin{eqnarray}\label{eq:example}
C_{\rm QCD}(y,x)&=&\sum_{q=u,d,s,\dots}\big<  \bar q(y)\Gamma^\prime q(y) \bar q(x)\Gamma q(x)\big>_{\rm QCD}
		\nonumber\\[-2mm]
&&\\[-2mm]
&=&-\sum_{q=u,d,s,\dots}\big< 
   \Tr\left\{ S_q(x,y)\Gamma^\prime S_q(y,x)\Gamma    \right\}\big>_{\rm QCD}  \nonumber \\
   &&+ \sum_{q=u,d,s,\dots}\big<    \Tr\left\{S_q(y,y)\Gamma^\prime\right\}
        \Tr\left\{S_q(x,x)\Gamma\right\}\big>_{\rm QCD}\,,\nonumber\
\end{eqnarray}
where $S_q$ is the propagator of the quark field $q$
 and  $\Gamma^{(\prime)}$ may contain Dirac- as well as colour-structures.
The trace is over Dirac- and colour-indices.
We will rewrite the correlation function $C_{\rm QCD}(y,x)$ as the sum of two new 
correlation functions: one being equivalent to the quark-connected 
contribution and the other being equivalent to the quark-disconnected 
contribution.
To this end for each quark $q$ we add a valence quark $q_v$ which is
mass degenerate to it, together with the corresponding ghost field.
The correlation $C_{\rm QCD}(y,x)$ can be rewritten as
\begin{eqnarray}\label{eqn:master}
C_{\rm QCD}(y,x)&=&\;\;\;\sum_{q=u,d,s,\dots}\big< \bar q(y)\Gamma^\prime q_v(y)
                \bar q_v(x)\Gamma q(x)\big>_{\rm PQQCD}
	  \\&&+\sum_{q=u,d,s,\dots}
        \big< \bar q(y)\Gamma^\prime q(y)
                \bar q_v^\prime(x)\Gamma q_v(x)
           \big>_{\rm PQQCD}\nonumber\\[0mm]
        &\equiv&C_{\rm PQQCD}^{\rm Conn}(y,x)+C_{\rm PQQCD}^{\rm Disc}(y,x)\nonumber\,.
\end{eqnarray}
The Wick contractions of the first correlator on the r.h.s. of the first 
equation lead to quark connected  diagrams
only, while those for the second correlator produce quark-disconnected
 diagrams only (see illustrations in figure \ref{fig:Wicks}). 
If we imagine for a moment having regularised the theory on the lattice it
is clear that the above equality holds non-perturbatively, since on each gauge
configuration the quark propagators and the fermionic determinant (and
therefore the weight of the configuration in the path integral) are the same in 
the two theories.
In the following 
we will show how $C_{\rm PQQCD}^{\rm Disc}(y,x)$ and $C_{\rm PQQCD}^{\rm Conn}(y,x)$ can be 
expressed in a suitable chiral effective theory, depending on the 
flavours entering the sum in Eq.~\ref{eq:example}.

\section{Hadronic vacuum polarisation}\label{sec:3}
The Euclidean hadronic vacuum polarisation (VP) tensor is defined as
\begin{equation}\label{eqn:pol_tensor}
\Pi_{\mu\nu}^{(N_f)}(q)=
	i\int d^4xe^{iqx}\langle
	J_\mu^{(N_f)}(x)J_\nu^{(N_f)}(0)
	\rangle = i C_{\mu\nu}(q)\,,
\end{equation}
where $J_\mu^{(N_f)}(x)=\sum\limits_{q=1}^{N_f}Q_q\bar q(x)\gamma_\mu q(x)$. For
$N_f=2$, $q=(u,d)$ and $Q_q=(2/3,-1/3)$ and for $N_f=2+1$,
	$q=(u,d,s)$ and $Q_q=(2/3,-1/3,-1/3)$.
Euclidean invariance and current conservation imply
\begin{equation}
\Pi_{\mu\nu}^{(N_f)}(q)=(q_\mu q_\nu-g_{\mu\nu}q^2)\Pi^{(N_f)}(q^2)\,.
\end{equation}
For space-like momenta, the relation between $\Pi_{\mu\nu}^{(N_f)}(q^2)$ and 
the lowest order hadronic contribution $a_\mu^{\rm HLO}$ 
to the anomalous magnetic moment of the 
muon has been derived in Ref.~\cite{Blum:2002ii,Gockeler:2003cw} and reads (suppressing the index $N_f$)
\begin{equation}
a_\mu^{\rm HLO}= \left( {{\alpha}\over{\pi}} \right)^2 \int_0^\infty dq^2 \, f(q^2)
\hat{\Pi}(q^2) 
\;,
\label{eq:amu}
\end{equation}
with
\be
f(q^2)={{m_\mu^2q^2Z^3(1-q^2Z)}\over{1+m_\mu^2q^2Z^2}}\;, \quad \quad
Z=-{{q^2-\sqrt{q^4+4m_\mu^2q^2 }}\over{2m_\mu^2q^2}}\,,
\ee
and $\hat{\Pi}(q^2)=4\pi^2\left[\Pi(q^2)-\Pi(0)\right]$.
We consider the iso-scalar meson two-point function $C_{\mu\nu}(q)$ in
Eq.~(\ref{eqn:pol_tensor}).
The Wick contractions for that correlator
lead to quark diagrams of type
(a) and (b) as illustrated in figure \ref{fig:Wicks}. 
By considering in some detail the two flavour
case we will first separate the disconnected contributions from the
connected ones in the way discussed in the previous section and then
set-up a computation in the resulting PQ$\chi$PT framework.

According to the discussion above we would need two valence quarks, one 
degenerate with the $u$ quark and one degenerate with the $d$ quark. 
However,
as we will always assume iso-spin to be an exact symmetry, it is enough to
introduce one valence quark, which we call $r$, with the corresponding 
ghost $r_{\rm g}$. We therefore have an $SU(3|1)$ chiral group.
The correlation in Eq.~(\ref{eqn:pol_tensor}) can then be decomposed as
\begin{eqnarray}
C^{}_{\mu\nu}(q)&\hspace{-2.5mm}=\hspace{-2.5mm}& \int d^4xe^{iqx}\bigg(
	{{4}\over{9}}\Big< j^{ur}_\mu(x) j^{ru}_\nu(0)\Big> + 
	{{1}\over{9}}\Big< j^{dr}_\mu(x) j^{rd}_\nu(0)\Big> +\nonumber\\[-1mm] 
\\[-1mm]
&&\qquad\qquad\;\;\;\;
	{{4}\over{9}}\Big< j^{uu}_\mu(x) j^{rr}_\nu(0)\Big> +
	{{1}\over{9}}\Big< j^{dd}_\mu(x) j^{rr}_\nu(0)\Big> -
	{{4}\over{9}}\Big< j^{uu}_\mu(x) j^{dd}_\nu(0)\Big> \bigg),\nonumber
\label{eq:su31splitting}
\end{eqnarray}
where $j^{q_1 q_2}_\mu(x)=\bar q_1(x) \gamma_\mu q_2(x)$.
The first two correlators on the r.h.s. are connected, whereas the last 
three represent the disconnected contributions to the hadronic VP tensor.
It is convenient to cast the quark-fields into a four-component vector 
$\psi^T=(u,d,r,r_{\rm g})$ and introduce the generators $T^a$, 
$a=1,\dots,15$ of the graded group $SU(3|1)$.
We use the conventions also employed in \cite{Giusti:2008vb},
\be
T^a=(T^a)^\dagger\;, \quad \quad \Str\left\{T^a\right\}=0\;, \quad
\Str\left\{T^aT^b\right\}={{1}\over{2}}g^{ab}\;, \quad a=1,\dots,15\,,
\ee
with the \textit{super-trace}
$\Str\left\{A\right\}=A_{11}+A_{22}+A_{33}-A_{44}$.
The matrix $g^{ab}$ reads
\begin{equation}
  g=\begin{pmatrix}1\cr
               &\ddots\cr
               &      &1\cr
               &      & &-\sigma_2\cr
               &      & &       &-\sigma_2\cr
               &      & &       &      &-\sigma_2\cr
               &      & &       &      &       &-1\cr\end{pmatrix}
  \begin{matrix}\left.\vphantom{\begin{matrix}1\cr
                                   &\ddots\cr
                                   &      &1\cr \end{matrix}}\right\}&
                                   \kern-1.5ex1-8\hfill\cr
          \left.\vphantom{\begin{matrix}-\tau^2\cr
                                         &\ddots\cr
                                         &      &-\tau^2\cr\end{matrix}}\right\}&
                                   \kern-1.5ex9-14\hfill\cr
          \left.\vphantom{\begin{matrix} 1\cr \end{matrix} }\right\}&
                                   \kern-1.5ex15\hfill\cr
         \end{matrix}
\end{equation}
where $\sigma_2$ is the second Pauli matrix. $T^1,\dots, T^8$ are the 
generators of the $SU(3)$ subgroup that acts on the sea and valence 
components, $T^9,\dots, T^{14}$ mix the quark with the ghost components,
$T^{15}$ is the diagonal matrix ${\rm diag}(1,1,1,3)/(2\sqrt{3})$. 
We also add $T^0$ with $\Str\{T^0\}=1/\sqrt{2}$
which is proportional to the unit matrix
in order to describe iso-scalar interactions.
For our choice of the generators $T^0,\,\dots,\,T^{15}$, 
the currents in Eq.~(\ref{eq:su31splitting}) can then be rewritten as
{\large\begin{equation}
\begin{array}{rcl}
j^{ur}_\mu(x)&=&\bar\psi(x)\gamma_\mu\left(T^4+i T^5\right)\psi(x)\,,\\[1mm]
j^{ru}_\mu(x)&=&\bar\psi(x)\gamma_\mu\left(T^4-i T^5\right)\psi(x)\,,\\[1mm]
j^{dr}_\mu(x)&=&\bar\psi(x)\gamma_\mu\left(T^6+i T^7\right)\psi(x)\,,\\[1mm]
j^{rd}_\mu(x)&=&\bar\psi(x)\gamma_\mu\left(T^6-i T^7\right)\psi(x)\,,\\[1mm]
j^{uu}_\mu(x)&=&\bar\psi(x)\gamma_\mu\left(
		\sqrt{2} T^0 - \frac{1}{\sqrt{3}} T^{15} + 
		 \frac{1}{\sqrt{3}} T^8 + T^3\right)\psi(x)\,,\\[2mm]
j^{dd}_\mu(x)&=&\bar\psi(x)\gamma_\mu\left(
		\sqrt{2} T^0 - \frac{1}{\sqrt{3}} T^{15} + 
		 \frac{1}{\sqrt{3}} T^8 - T^3\right)\psi(x)\,,\\[2mm]
j^{rr}_\mu(x)&=&\bar\psi(x)\gamma_\mu\left(
		\sqrt{2} T^0 - \frac{1}{\sqrt{3}} T^{15} - 
		 \frac{2}{\sqrt{3}} T^8\right)\psi(x)\,.
\end{array}
\end{equation}}

\noindent This form is more suited for the PQ$\chi$PT computation, as it will 
become clear in the following.

Before concluding this section we note that by using again iso-spin 
symmetry we could have separated the disconnected and connected parts
in the correlator above without introducing any additional valence quark.
We have used this completely equivalent approach in Ref.~\cite{Juttner:2009yb}.
Here however, we preferred to introduce a graded flavour group
to provide an example
for how one has to proceed for the the more 
general but also more complicated case of $2+1$ flavours.

\section{PQ$\chi$PT for the connected and disconnected parts of the hadronic VP}
In this section we briefly introduce those parts of the (partially quenched)
chiral Lagrangian up to $O(p^4)$ which contribute to the connected as well
as to the disconnected piece of the VP \cite{Gasser:1984gg,Gasser:1983yg}. We took care that
the discussion applies to any choice for the (graded) symmetry group. In the 
next section we will then present results for the cases of 
$SU(3|1)$, $SU(4|2)$, and $SU(4|1)$ flavour groups. These are the relevant 
symmetry groups for the description of the contributions to the VP in the $N_f=2$-theory without and with a quenched
strange quark and for the $N_f=2+1$-theory, respectively.
\subsection{$O(p^2)$-Lagrangian}
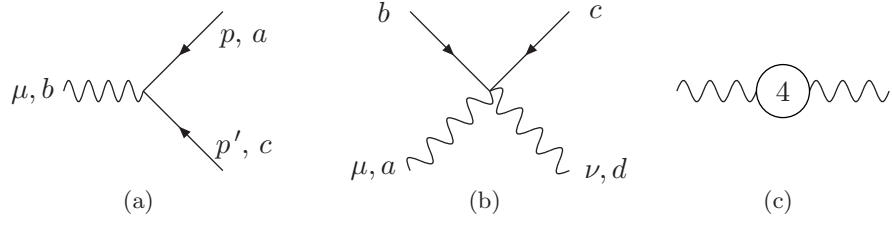
\begin{figure}
\centering
\subfigure[]{
 \begin{picture}(120,60)(-60,-30)
  \Photon(-30,0)(00,0){4}{4}
  \Text(-42,  0)[c]{$\mu, b$}
  \Text(+38, 20)[c]{$p,\,a$}
  \Text(+38,-20)[c]{$p^{\,\prime},\,c$}
  \ArrowLine(30,30)(0,0)
  \ArrowLine(30,-30)(0,0)
 \end{picture}
}
\subfigure[]{
 \begin{picture}(120,60)(-60,-30)
  \Photon(-30,-30)(00,0){4}{4}
  \Photon(0,0)(30,-30){4}{4}
  \ArrowLine(-30,30)(0,0)
  \ArrowLine(30,30)(0,0)
  \Text(-44,-30)[c]{$\mu,a$}
  \Text(+44,-30)[c]{$\nu,d$}
  \Text(-40,+30)[c]{$b$}
  \Text(+40,+30)[c]{$c$}
 \end{picture}
}
\subfigure[]{
  \label{fig:feyn_c}
  \begin{picture}(80,40)(-40,-30)
	\Photon(-40,0)(-10,0){4}{3}
	\Photon(10,0)(40,0){4}{3}
      \GCirc(0,0){10}{1}
      \Text(0,0)[c]{4}
  \end{picture}
}
\caption{Contributing vertices in the effective theory.}\label{fig:vertices}
\end{figure}

For a generic graded flavour group, the leading order chiral Lagrangian is 
\cite{Gasser:1984gg,Gasser:1983yg,Bernard:1992mk,Bernard:1993sv,Sharpe:2000bc}
\begin{equation}
\mathcal{L}^{(2)}=\frac{F^2}{4}\Str\left\{D_\mu U D_\mu U^\dagger\right\}
	-\frac 12 B F^2 \Str\left\{M U^\dagger + M^\dagger U\right\}\,,
\end{equation}
where $D_\mu U=\partial_\mu U+i v_\mu U -i U v_\mu$ provides the
coupling of the meson field to an external vector source $v_\mu$
and where
\begin{equation}
M={\rm diag}\left(m_1,\dots,m_{N_f},
	m_1^{\rm valence},\dots,m_{N}^{\rm valence},
	m_1^{\rm ghost},\dots,m_{N}^{\rm ghost} \right)\,,
\end{equation} 
contains the dynamical, valence-, and ghost-quark masses.
By chirally expanding\linebreak
$U=\exp\left(2i\frac {\phi^a T^a} F\right)$, where $T^a$ are the 
$\tilde N=(N_f+2N)^2-1$ generators of the corresponding 
$SU(N_f+N|N)$ graded symmetry group 
(with the generators normalised as in the previous section) 
and where the $\phi^a$ are the Goldstone-boson/fermion fields,
we arrive at the $O(p^2)$-expression  
\begin{equation}\label{eqn:O2Lag}
 \mathcal{L}^{(2)}=
	\frac 12 \,g^{ab} \partial_\mu \phi^a\partial_\mu \phi^b
	+ 2 B \tilde M^{ab}\phi^a \phi^b+\mathcal{L}^{(2)}_{\rm int}\,,
\end{equation}
where $\tilde M^{ab}=\Str\left\{MT^a T^b\right\}$. With 
$v_\mu=v_\mu^{a}{T^a}$, the part of the interaction Lagrangian that
is of relevance here is
\begin{equation}	
\mathcal{L}^{(2)}_{\rm int}=
	\underbrace{
	-\sum\limits_{k=1}^{\tilde N}
		C_k^{\;bc}g^{ak}\partial_\mu \phi^a v_{\mu}^{b}\phi^c
		}_{a)}
	\underbrace{
	+\frac 12 
	\sum\limits_{k=1}^{\tilde N}\sum\limits_{l=1}^{\tilde N}
	C_k^{\;ab}C_l^{\;cd}g^{kl}v_{\mu}^a \phi^b\phi^c v_{\mu}^{d}}_{b)}\,,
\end{equation}
where
\begin{equation}
C_a^{\;bc}= -2 i \sum\limits_{k=1}^{\tilde N}
	 \Str \left\{[T^b,T^c]\,T^k \right\}g^{k a}\,,
\end{equation}
are the structure constants of the underlying (graded) 
symmetry group with the\linebreak (anti-)commutator 
\begin{equation}
[T^a,T^b]\equiv T^a T^b-(-)^{\eta_a\,\eta_b}\,T^b T^a\,.
\end{equation}
The $\eta_a$ (and correspondingly for $\eta_b$) 
are  1 if $T^a$ mixes valence or sea
quarks with a ghost and 0 otherwise (cf.~\cite{Weinberg3}).
The Feynman-rules for the vertices illustrated in figure
\ref{fig:vertices} (a) and (b) are then determined as
\begin{equation}
\begin{array}{cl}
	a)& 	\qquad\qquad\;\;
	-\Big( 
		\tilde C_a^{\;bc}\,p_\mu+
		\tilde C_c^{\;ba}\,p^{\,\prime}_\mu
	\Big)\,,\\
	b)& 	i\eta_{\mu\nu}\sum\limits_{k,l=1}^{\tilde N}g^{kl}
	\Big( C_{k}^{\;ab}C_{l}^{\;cd}+ C_{k}^{\;ac}C_{l}^{\;bd}\Big)\,,
\end{array}
\end{equation}
with $\tilde C_a^{\;\;bc}= C_i^{\;\,bc}g^{ai}$.

\subsection{$O(p^4)$-Lagrangian}
The relevant terms of the $O(p^4)$ Lagrangian \cite{Gasser:1984gg,Gasser:1983yg,Kaiser:2000ck,Kaiser:2000gs}, i.e. those parts that 
have a non-zero matrix element between single external vector-sources, are 
\begin{equation}\label{eqn:Lcounter}
\begin{array}{rcl}
\mathcal{L}^{(4)}&=&
	X_1\,\Str\{ \hat v_{\mu\nu} \hat v_{\mu\nu}\}
	 + X_2\,\Str\{ v_{\mu\nu} \} \Str  \{v_{\mu\nu} \}\,,
\end{array}
\end{equation}
with 
\begin{equation}
v_{\mu\nu}=\partial_\mu v_\nu -\partial_\nu v_\mu\,,
\end{equation}
and where $\hat v_{\mu\nu}$ is the trace-less part 
$v_{\mu\nu}-\frac 1{N_f}\,\Str\{{v_{\mu\nu}}\}$.
As summarised in table
\ref{tab:GLcoeffs} the coefficient $X_{1}$ is the shorthand 
notation for the Gasser-Leutwyler
low energy constants of the underlying symmetry group (see for example
\cite{Gasser:1983yg,Gasser:1984gg,Bijnens:1999hw}) and $X_2$ 
 parameterises the effective dynamics of flavour-diagonal
contributions.
\begin{table}
\begin{center}
\begin{tabular}{lcclllll}
\hline\hline\\[-4mm]
		&$X_1$	&$X_2$	\\
\hline&&&&\\[-4mm]
$\SUtwo$, $SU(3|1)$, $SU(4|2)$   &$-4h_2$	&$h_s$  \\
$\SUthree$, $SU(4|1)$ 		&$L_{10}+2H_1$	&$H_s$			\\[1mm]
\hline\hline
\end{tabular}\caption{Gasser-Leutwyler coefficients.}\label{tab:GLcoeffs}
\end{center}
\end{table}
The vertices corresponding to these Lagrangians have the following form
(cf. figure \ref{fig:vertices} (c)):
\begin{equation}\label{eqn:counter:SU2}
\begin{array}{cll}
 c)& i\,2\,X_1\,g^{ab}
	\left(\eta_{\mu\nu}p^2-p_\mu p_\nu \right)\, &{\rm for\,}a,b>0\,,\\
 c)& i\,2\, X_2 \,\;\;\;\;\;\left(\eta_{\mu\nu}p^2-p_\mu p_\nu \right)\,& {\rm for\,}a=b=0\,.
\end{array}
\end{equation}
\subsection{Contributing diagrams}
\begin{figure}
	\centering
   \subfigure[]{
    \begin{picture}(80,40)(-40,-11)
  	\Photon(-40,0)(-20,0){4}{2}
  	\Photon(20,0)(40,0){4}{2}
	\GOval(0,0)(15,20)(0){1}
	\GCirc(-20,0){4}{.3}
	\GCirc(20,0){4}{.3}
	\Text(-45,0)[c]{a}
	\Text(+45,0)[c]{b}
    \end{picture}
    \label{fig:feyn_a}
   }\hspace{8mm}
   \subfigure[]{
    \label{fig:feyn_b}
    \begin{picture}(80,40)(-40,-11)
  	\Photon(-40,0)(40,0){4}{8}
	\GOval(0,20)(20,10)(00){1}
	\GCirc(0,0){4}{.3}
	\Text(-45,0)[c]{a}
	\Text(+45,0)[c]{b}
    \end{picture}
   }
   \hspace{8mm}
   \subfigure[]{
    \label{fig:feyn_c2}
    \begin{picture}(80,40)(-40,-11)
  	\Photon(-40,0)(-10,0){4}{3}
  	\Photon(10,0)(40,0){4}{3}
	\GCirc(0,0){10}{1}
	\Text(0,0)[c]{4}
	\Text(-45,0)[c]{a}
	\Text(+45,0)[c]{b}
    \end{picture}
   }
 \caption{Feynman diagrams contributing to the hadronic VP at 
1-loop level: (a) unitary, (b)
	tadpole, (c) $O(p^4)$-insertion}
 \label{fig:FMdiags}
\end{figure}
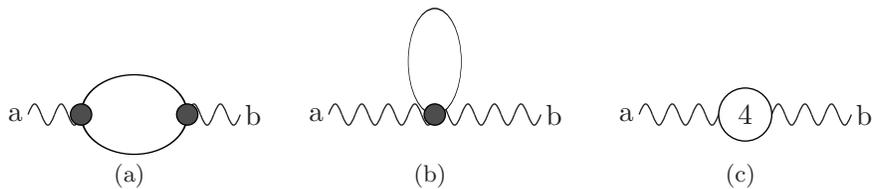

At NLO only the diagrams in figure
\ref{fig:FMdiags}~(a) and \ref{fig:FMdiags}~(b) 
contribute dynamically to the VP and the diagram 
in figure~\ref{fig:FMdiags}~(c) provides counter terms. 
Let $\psi$ be a flavour vector of $N_f$ dynamical quarks, $N$ partially
quenched valence quarks and $N$ corresponding ghost-quarks.
The expression in the effective theory for 
the Fourier transform of the correlator
$i\langle \bar\psi(x) \gamma_\mu T^a\psi(x)\, \bar\psi(0) \gamma_\nu T^b\psi(0)\rangle$ 
at NLO in chiral perturbation theory is of the form
\begin{equation}
\Pi_{\mu\nu}^{(a,b)}(q,\mu)=\Pi_{\mu\nu}^{(a,b),\rm unit.}(q)
			+\Pi_{\mu\nu}^{(a,b),\rm tadp.}(q)
			+\Pi_{\mu\nu}^{(a,b),\rm count.}(\mu)\,,
\end{equation}
where $\mu$ is the renormalisation scale.
The first term on the r.h.s. corresponds to the unitary diagram,
\begin{eqnarray}\label{eq:unitary}
\displaystyle
\Pi_{\mu\nu}^{(a,b),\rm unit.}(q)&=&-\frac i2\,
	\tilde C_i^{\;aj}\,\tilde C_k^{\;bl}\\
&&\hspace{-1cm}\times		\int\frac{d^d k}{(2\pi)^d}
	\, (q+2k)_\mu(q+2k)_\nu\, 
	G^{ik}\big(k^2\big)\,G^{jl}\big((q-k)^2\big)\nonumber\,,
\end{eqnarray}
where the overall
factor $\frac 12$ is a 
symmetry factor. 
For the second term, the tadpole-diagram (b), one derives
\begin{eqnarray}\label{eq:tadpole}
\Pi_{\mu\nu}^{(a,b)\,\rm tadp.}
	&=&i\,\eta_{\mu\nu}\,C_i^{\;aj}\,\tilde C_{i}^{\;kb}
	\int\frac{d^d k}{(2\pi)^d}\,
	G^{jk}(k^2)\,.
\end{eqnarray}
For the  $SU(3|1)$ theory the 
meson propagators $G^{ab}(k^2)$ can be found 
e.g. in Ref.~\cite{Giusti:2008vb} and we provide the corresponding
expressions for $SU(4|1)$ and $SU(4|2)$ in appendix \ref{app:props}. 
The last term, $\Pi_{\mu\nu}^{(a,b),\rm count.}(\mu)$,
contains the counter terms. We compute $\Pi_{\mu\nu}^{(a,b)}(q,\mu)$
in dimensional regularisation as
explained in detail in Ref.~\cite{Golowich:1995kd} where also the solutions
for the loop integrals have been derived and reduced to the
integral $\bar B_{21}(q^2,M^2)$ defined in appendix \ref{app:B21}.

\section{Results and applications to Lattice QCD}
Here we present the results for the VP at NLO in chiral
perturbation theory for various underlying graded flavour-symmetry groups. In
particular, we consider the theories with  $N_f=2$ flavours without or with
an additional quenched strange quark and also the $N_f=2+1$ theory.
In each case we present the expressions in the effective theory for the full 
VP and also for the contributions from quark-connected and
quark-disconnected diagrams.
\subsection{$N_f=2$ in $SU(3|1)$ PQ$\chi$PT}
For the theory with two dynamical and degenerate light quarks only we found the following expressions for the VP:
\begin{equation}
\begin{array}{l@{\hspace{1mm}}c@{\hspace{1mm}}c@{\hspace{0mm}}l@{\hspace{0cm}}l@{\hspace{1mm}}c@{\hspace{1mm}}lll}
 \Pi_{\rm Full}^{(3|1)}(q^2)&=-&&\Big(\Lambda^{(3|1)}(\mu)
	&+\,\frac{2}{9}h_s&+&i4\bar B_{21}(\mu^2,q^2,M^2_\pi)\Big)\,,\\[3mm]
 \Pi_{\rm Conn}^{(3|1)}(q^2)&=-&\frac{10}{9}&\Big(\Lambda^{(3|1)}(\mu)&&+&i 4\bar B_{21}(\mu^2,q^2,M_\pi^2)\Big)\,,\\[3mm]
 \Pi_{\rm Disc}^{(3|1)}(q^2)&=\phantom{-}&\frac{1}{9}&\Big(\Lambda^{(3|1)}(\mu)&-\,{2}h_s&+&i4\bar B_{21}(\mu^2,q^2,M_\pi^2)\Big)\,,
\end{array}
\end{equation}
where
\begin{equation}
\begin{array}{l@{\hspace{1mm}}c@{\hspace{1mm}}l@{\hspace{1mm}}c@{\hspace{1mm}}r@{\hspace{1mm}}c@{\hspace{1mm}}r@{\hspace{1mm}}ccc}
\Lambda^{(3|1)}(\mu)&=&-8 h_2(\mu)\,.\\[3mm]
\end{array}
\end{equation}
The integral $\bar B_{21}(\mu^2,q^2,M^2)$ is defined in appendix 
\ref{app:B21}. 
All contributions are parameterised in terms of the low-energy constant
$h_2$ and the pion mass $M_\pi$. 
The full expression as well as the disconnected piece also depend
on the parameter $h_s$. This is 
a peculiarity of the two-flavour theory and we will see in the next
sub-sections that the corresponding expressions in the presence of a 
strange valence quark do not depend on this additional parameter.
Hence, a prediction of the quark-disconnected
diagram is not possible without the knowledge of $h_2$
and $h_s$. These parameters however,
can in principle be determined from simulations of lattice QCD and $h_s$ can 
also be obtained by matching to the expressions in the next section.

The application we have in mind when discussing the VP is the
leading hadronic contribution to the muon anomalous magnetic moment defined
in Eq.~(\ref{eq:amu}). As discussed in section \ref{sec:3}, 
$a_\mu^{\rm HLO}$ depends on 
$\hat{\Pi}(q^2)=4\pi^2\left[\Pi(q^2)-\Pi(0)\right]$. We observe that
all reference to the low-energy constants disappears in this difference
at NLO in the effective theory and we find \cite{Juttner:2009yb}
\begin{equation}
{{\hat{\Pi}_{\rm Disc}(q^2)}\over{\hat{\Pi}_{\rm Conn}(q^2)}}=-{{1}\over{10}}\,.
\end{equation}
At this order in the effective theory it is therefore sufficient to
compute the quark-connected piece contributing to $a_\mu^{\rm HLO}$ and
then correct for the quark-disconnected piece using the above relation which
predicts a 10\% negative shift for all values of the momentum (note in this
context the discussion in section \ref{subsec:disclaimer}).
The disconnected contribution at NLO turns out to have the same
momentum and quark-mass dependence as the connected piece. 
\subsection{$N_f=2$ plus a quenched strange quark in $SU(4|2)$ PQ$\chi$PT}\label{subsec:4bar2}
\begin{figure}
	\centering
	\subfigure[]{
	\begin{minipage}{.45\linewidth}
	\psfrag{xlabel}[t][t][1][0]{$q^2/{\rm GeV^2}$}
	\psfrag{ylabel}[c][t][1][0]{
		$\hat\Pi_{\rm Disc}(q^2)/\hat\Pi_{\rm Conn}(q^2)$}
	\hspace{-15mm}\epsfig{scale=.9,file=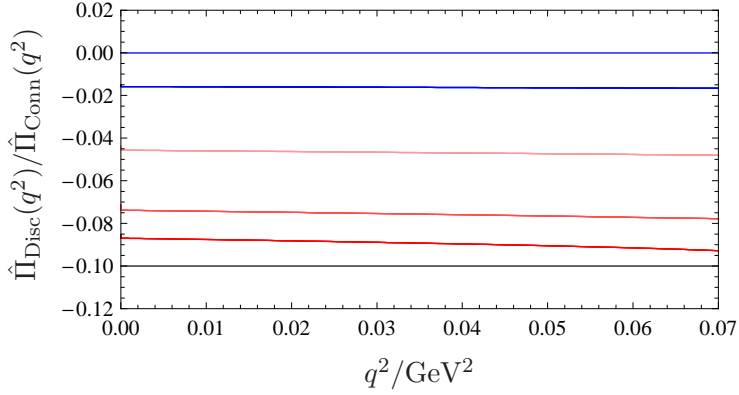}\\[2mm]
	\end{minipage}
	}\\
	\subfigure[]{
	\begin{minipage}{.45\linewidth}
	\psfrag{xlabel}[t][t][1][0]{$q^2/{\rm GeV^2}$}
	\psfrag{ylabel}[c][t][1][0]{
		$\hat\Pi_{\rm Disc}(q^2)/\hat\Pi_{\rm Conn}(q^2)$}
	\hspace{-15mm}\epsfig{scale=.9,file=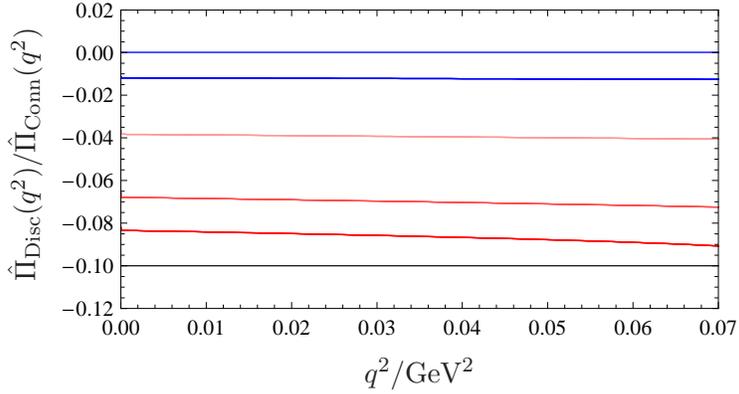}\\[2mm]
	\end{minipage}
	}
	\caption{The plots illustrate the ratio of the contributions
		from quark-disconnected and quark-connected diagrams
		to the VP as a function of the momentum $q^2$ 
		at NLO in chiral perturbation theory for the cases (a) 
		$N_f=2$ with a quenched strange quark,
		(b) $N_f=2+1$. In both cases the lines are from top to bottom:
		the limit $M_\pi =M_K$, then fixed $M_K=495$MeV and
		$M_\pi=400,\,300,\,200,\,139$MeV. The bottom most-line
		at -1/10 is the result for the $N_f=2$ theory.
		}
	\label{fig:results}
\end{figure}

For the theory with two dynamical and degenerate light quarks and one 
quenched strange quark one needs to consider  $SU(4|2)$ PQ$\chi$PT, 
where we found the following expressions for the vacuum
polarisation: 
\begin{equation}
\begin{array}{l@{\hspace{0mm}}c@{\hspace{0mm}}c@{\hspace{0mm}}c@{\hspace{0mm}}r@{\hspace{1mm}}c@{\hspace{1mm}}r@{\hspace{1mm}}c@{\hspace{1mm}}c@{\hspace{0mm}}c}
\Pi_{\rm Full}^{(4|2)}(q^2)&=&\Lambda^{(4|2)}(\mu)&-4i\Big(&
		         \bar B_{21}(\mu^2,q^2,M_\pi^2)
		&-&\frac 19 \bar B_{21}(\mu^2,q^2,M_{ss}^2)
		&+&\frac {4}9 \bar B_{21}(\mu^2,q^2,M_{K}^2)&\Big)\,,\\[5mm]
\Pi^{(4|2)}_{\rm Conn}(q^2)&=&\Lambda^{(4|2)}(\mu)	&-4i\Big(&
		\frac{10}{9}\bar B_{21}(\mu^2,q^2,M_\pi^2)
		&&&+&\frac{2 }{9}\bar B_{21}(\mu^2,q^2,M_{K}^2)&\Big)\,,\\[5mm]
\Pi^{(4|2)}_{\rm Disc}(q^2)&=&&-4i\Big(&
		-\frac{1}{9}\bar B_{21}(\mu^2,q^2,M_\pi^2)
		&-&\frac{1 }{9}\bar B_{21}(\mu^2,q^2,M_{ss}^2)
		&+&\frac{2 }{9}\bar B_{21}(\mu^2,q^2,M_{K}^2)&\Big)\,,
\end{array}
\end{equation}
where
\begin{equation}
\begin{array}{l@{\hspace{1mm}}c@{\hspace{1mm}}c@{\hspace{1mm}}c@{\hspace{1mm}}r@{\hspace{1mm}}c@{\hspace{1mm}}r@{\hspace{1mm}}ccc}
\Lambda^{(4|2)}(\mu)&=&8h_2(\mu)\,.
\end{array}
\end{equation}
The full expression and the connected piece depend on the low energy constant
$h_2$ and the pion and kaon mass. An additional dependence on the 
mass of the strange-quark enters in terms of the mass 
$M^2_{ss}=2B m_s$ of an 
un-physical meson made by the  two quenched
strange quarks.
It is well known that in the $N_f=2+1$-theory 
the quark-disconnected contribution to the
hadronic VP vanishes in the limit of equal quark-masses.
This $SU(3)$
symmetry prohibits the presence of low-energy constants in the expression
for the quark-disconnected diagrams at NLO and this is indeed what we found
when computing $\Pi^{(4|2)}_{\rm Disc}(q^2)$. At NLO in the effective theory
we therefore 
provide an entirely parameter-free prediction of the quark-disconnected 
diagram. 
Moreover, as in the case of the $N_f=2$-theory, also $\hat{\Pi}(q^2)$ is free of 
low-energy constants and a
parameter-free prediction for the ratio of the quark-disconnected 
contribution to the quark-connected contribution can be made. 
As can be seen in figure \ref{fig:results} (a),
in the limits $M_K\to\infty$ and $M_K=M_\pi$, the above formulae reproduce the 
results for the 
$N_f=2$ theory and the vanishing of the disconnected piece in the
$M_K=M_\pi$-limit, respectively. 
\subsection{$N_f=2+1$ in $SU(4|1)$ PQ$\chi$PT}\label{subsec:4bar1}
For the theory with two dynamical and degenerate light quarks and one dynamical
strange quark we found the following expressions for the vacuum
polarisation: 
\begin{equation}
\begin{array}{l@{\hspace{0mm}}c@{\hspace{0mm}}c@{\hspace{0mm}}c@{\hspace{0mm}}r@{\hspace{1mm}}c@{\hspace{1mm}}r@{\hspace{1mm}}c@{\hspace{1mm}}c@{\hspace{0mm}}c}
\Pi^{(4|1)}_{\rm Full}(q^2)&=&\Lambda^{(4|1)}(\mu)&-4i\Big(&
		         \bar B_{21}(\mu^2,q^2,M_\pi^2)
		&&&+& \bar B_{21}(\mu^2,q^2,M_{K}^2)&\Big)\,,\\[5mm]
\Pi^{(4|1)}_{\rm Conn}(q^2)&=&\Lambda^{(4|1)}(\mu)	&-4i\Big(&
		\frac{10}{9}\bar B_{21}(\mu^2,q^2,M_\pi^2)
		&+&\frac{1 }{9}\bar B_{21}(\mu^2,q^2,M_{ss}^2)
		&+&\frac{7}{9}\bar B_{21}(\mu^2,q^2,M_{K}^2)&\Big)\,,\\[5mm]
\Pi^{(4|1)}_{\rm Disc}(q^2)&=&&-4i\Big(&
		-\frac{1}{9}\bar B_{21}(\mu^2,q^2,M_\pi^2)
		&-&\frac{1 }{9}\bar B_{21}(\mu^2,q^2,M_{ss}^2)
		&+&\frac{2 }{9}\bar B_{21}(\mu^2,q^2,M_{K}^2)&\Big)\,,
\end{array}
\end{equation}
where
\begin{equation}
\begin{array}{l@{\hspace{1mm}}c@{\hspace{1mm}}l@{\hspace{1mm}}c@{\hspace{1mm}}r@{\hspace{1mm}}c@{\hspace{1mm}}r@{\hspace{1mm}}ccc}
\Lambda^{(4|1)}&=&-2\left( L_{10}(\mu)+2 H_1(\mu)\right)\,.\\[3mm]
\end{array}
\end{equation}
As in the previous sub-section,
the full expression and the connected piece depend on low-energy constants,
this time
$L_{10}$ and $H_1$, and the pion and kaon mass. Also here 
a dependence on the 
mass of the strange-quark enters in terms of the mass 
$M^2_{ss}=2B m_s$ of an 
un-physical meson made of a dynamical and a quenched
strange quark.
In the same way as in the previous section $SU(3)$
symmetry prohibits the presence of low-energy constants in the expression
for the quark-disconnected diagram at NLO.
At this order in the effective theory we are therefore able to 
provide an entirely parameter-free prediction of the quark-disconnected 
diagram. 
Again, $\hat{\Pi}(q^2)$ is free of 
low-energy constants and a
parameter-free prediction for the ratio of the quark-disconnected 
contribution to the quark-connected contribution can be made. 
We illustrate this result in figure \ref{fig:results} (b).
In the same way as for the $N_f=2$ theory with a quenched strange quark,
in the limits limits $M_K\to\infty$ and $M_K=M_\pi$, the above formulae reproduce
the results for the $N_f=2$ theory and the vanishing of the disconnected piece in the
$SU(3)$-limit, respectively. As a further check, our formulae reproduce 
the result in Ref.~\cite{Golowich:1995kd}.
From the comparison 
of the plots (a) and (b) we see little influence of the dynamical strange quark.
\subsection{Discussion}\label{subsec:disclaimer}
Before we suggest applications to simulations of lattice QCD we would like
to point out obvious
limitations of the above formulae. While the underlying idea of the approach, 
i.e. Eq.~(\ref{eqn:master}), is exact in QCD and in the chiral effective theory,
the results presented
here are only valid up to NLO chiral perturbation theory.
If at all, this approximation only holds for small quark-masses and only
for small values of the momentum. There are 
in fact doubts that the 
physical strange quark mass can be described reliably within chiral perturbation
theory \cite{Allton:2008pn}. One might hope that cancellations in the 
ratio $\hat\Pi_{\rm Disc}(q^2)/\hat\Pi_{\rm Conn}(q^2)$ will improve the
 convergence
of the expansion in the meson masses and the momentum. However, 
only expensive numerical simulations in lattice QCD and higher order
computations in the effective theory will be able to tell.

At least as important is the fact that instead of 
coupling to two pions, the photon
will also couple to vector resonances. It is \textit{a priori} not clear 
if at all or in which kinematical regime chiral perturbation theory 
parameterises
these dynamics correctly 
in terms of the low-energy constants of the chiral effective
Lagrangian. Aubin and Blum \cite{Aubin:2006xv} for example found 
indications for vector-dominance in the analysis of their lattice QCD 
data for the hadronic VP.
We  expect sizeable corrections to the expressions found here.
There do exist models, generally known as ``resonace $\chi$PT `` models,
where the vector resonances are
dynamical degrees of freedom (see e.g. \cite{Ecker:1988te}). 
Conceptually these models are not as solid as $\chi$PT, to the extent that
no systematic power counting can be formulated.
While clearly desirable, 
a study of how the vector degrees of freedom will modify the expressions 
is beyond the scope of the current work but in principle 
technically straight forward within the framework of resonace $\chi$PT,
which is however subject to the theoretical issues mentioned above. 

Given these remarks, the results presented here can only provide an 
order-of-magnitude estimate and should be used with great care.
In the absence of 
any other quantitative knowledge of the quark-disconnected diagrams, the 
results can be used in order to correct the lattice
data for the connected contributions.
\subsection{Applications to lattice QCD}\label{subsec:lat-app}
{\bf Partially twisted boundary conditions for the VP:} Partially twisted 
boundary conditions \cite{Sachrajda:2004mi,Bedaque:2004ax,Bedaque:2004kc,deDivitiis:2004kq,Tiburzi:2005hg,Flynn:2005in} by now have become a standard
tool in lattice hadron phenomenology. For the boundary condition
 $q_{i}(x_k+L)=e^{i\theta_{k,i}}q_{i}(x)$ 
for the valence quark-flavours $q_1$ and $q_2$ 
of a pseudo-scalar meson of mass $m$ in a finite lattice 
of spatial extent $L$, the 
dispersion relation takes the form
$E({\vec\theta}_1,{\vec\theta}_2)=\sqrt{m^2+({\vec \theta}_1/L-{\vec \theta}_2/L)^2}$  \cite{deDivitiis:2004kq,Sachrajda:2004mi,Flynn:2005in},
which shows how the periodicity of the fermionic fields can be modified
in order to induce spacial momentum to hadrons. 
Besides an exponentially suppressed and computable finite size effect
the choice of quark-field 
boundary conditions for processes with only one initial and/or
final hadron state introduces no further systematics
\cite{Sachrajda:2004mi}.
The effect of partial twisting cancels however
in flavour-neutral mesons, as is clear from the form of the above
dispersion relation. Due to the flavour-diagonal
structure of the electro-magnetic current, 
partial twisting would therefore naively have
no effect on the hadronic VP. The 
prescription introduced here however,
allows for isolating a flavour-off-diagonal connected contribution
and thus for inducing arbitrary values of the momentum
for the VP. In other words we can assign different twisting angles 
to the quarks $q$ and $q_v$ in Eq.~\ref{eqn:master}, which has the net
effect of inducing momentum in the connected part.
The effectiveness of this method in the numerical computation of the connected
part of the hadronic VP will be illustrated in a forthcoming 
publication~\cite{Mainz:g-2}.
The relations in the chiral effective theory between the quark-connected
and the quark-disconnected piece then allow to predict also the 
momentum-dependence of the disconnected piece. 
A very similar argument was already used
in Refs.~\cite{Boyle:2007wg,Jiang:2008te} in order to justify the use of 
partially twisted boundary conditions for the pion's vector form factor.
The exponential suppression in the volume of the isospin breaking 
introduced by partially twisting only one of the light quarks is also 
discussed in Refs.~\cite{Boyle:2007wg,Jiang:2008te}.

\section{Conclusions and outlook}
In this paper we show how  quark-disconnected
contributions to hadronic $n$-point functions can be estimated
within the framework of 
partially quenched chiral perturbation theory. As an example we derive predictions
for the quark-disconnected contribution to the photon's 
vacuum polarisation for QCD with two degenerate light quarks,
with and without a quenched strange quark, and also for the case of 
the three-flavour theory. In the presence of a quenched strange quark 
the prediction for the quark-disconnected contribution turns out to be 
parameter-free. 

The vacuum polarisation is currently computed by
various lattice collaborations with the aim to make first-principles predictions
for the leading hadronic contribution to the muon anomalous moment 
\cite{Aubin:2006xv,Renner:2009by,Juttner:2009yb}. Since  
precise computation of quark-disconnected correlators from first
principles
still remain numerically extremely expensive they are often neglected in 
such computations. The 
approach presented here will allow to improve estimates of the systematic
error introduced in this way.

The method can be applied to other physical observables of interest that
receive contributions from quark-disconnected diagrams, like
the strange-quark
content in the nucleon, the nucleon electric dipole moment 
 or the pion scalar form-factor. 
The discussion presented here is limited to NLO chiral perturbation theory. 
As a next step we plan to include the vector degrees of 
freedom and it would also be important to
extend the computation for the case of the vacuum polarisation to higher 
orders in the chiral expansion 
in order to assess how robust the predictions reported here are.
The inclusion of the leading lattice artifacts in the effective theory
is also straight forward within the framework set up by
Sharpe and Singleton in Ref.~\cite{Sharpe:1998xm}.

A spin-off of the ideas presented here is a method that allows for projecting
(the connected part of) correlators containing flavour-diagonal currents 
onto any desired momentum using partially twisted boundary 
conditions.\\[2mm]

\vspace{-2mm}
\noindent{\bf{Acknowledgments:}} We thank Dalibor Djukanovic, Silvia Necco, Hartmut Wittig, and Steve Sharpe for illuminating discussions.

\vspace{.8cm}
\newpage\appendix{\noindent{\bf{\large Appendices}}}
\section{Propagators in chiral perturbation theory}\label{app:props}
Here we provide the definition of propagators in PQ$\chi$PT
with the graded symmetry groups $SU(4|1)$ and $SU(4|2)$.
For simplicity all results are quoted in the 
limit where the Goldstone bosons and the corresponding ghosts are 
mass-degenerate. This limit
is sufficient for the discussion in the main body of this paper.
For completeness
we note that the case $SU(3|1)$ has been worked out nicely
in \cite{Giusti:2008vb}
and for most of the discussion we adopt the conventions laid out there.
\subsection{Propagators in $SU(4|1)$ chiral perturbation theory}
We consider the case of the effective theory with $N_f=2+1$ dynamical 
flavours with two degenerate  
light quarks of mass $m$, one strange quark
of mass $m_s$ and one quenched $r$ quark with 
$m_s=m_r= m_{r_g}$.
The dynamical (not necessarily physical) 
degrees of freedom in the effective theory have masses
\begin{equation}
M^2_\pi=2 B m\,,\qquad M^2_K=B(m+m_s)\,,\qquad
	M^2_{ss}=2 B m_s\,.
\end{equation}
The propagator $G^{ij}(k^2)$ is defined as follows:
The Lagrangian in Eq.~(\ref{eqn:O2Lag})  with 
$\mathcal{L}_{\rm int}^{(2)}=0$ can be rewritten as
\begin{equation}
\mathcal{L}^{(2)}=
\frac 12\, g^{ab}\left(
	\partial_\mu \phi^a \partial_\mu \phi^b+M_a^2\phi^a\phi^b\right)+
	\frac{1}{6}h^{ab}(M_\pi^2-M_K^2) \phi^a \phi^b\,,
\end{equation}
where
\begin{equation}
M_a^2=\left\{\begin{array}{lll}
M_\pi^2&{\rm for}&a=1,2,3\,,\\
\frac 12\left(M_\pi^2+M_K^2\right)&{\rm for}&a=4,5,6,7,9,10,11,12,16,17,18,19\,,\\
M_{ss}^2&{\rm for}&a=8,13,14,15,20,21,22,23,24\,,\\
\end{array}\right.
\end{equation}
and where for our choice of the $SU(4|1)$-generators
\begin{equation}
\begin{array}{lcl}
h^{ab}&=&\;\;\;(\sqrt{2}\,g^{a\,8}+g^{a\,15}-g^{a\,24})(\sqrt{2}\,g^{b\,8}+g^{b\,15}-g^{b\,24})\,,\\[1mm]
l^{ab}&=&-(\sqrt{2}\,g^{a\,8}+g^{a\,15}+g^{a\,24})(\sqrt{2}\,g^{b\,8}+g^{b\,15}+g^{b\,24})\,.
\end{array}
\end{equation}
In terms of these matrices the propagator is given by
\begin{eqnarray}
G^{ab}(k^2)&=& 
	g^{ab}G_1(k^2,M_a^2)
	+\frac{1}{3}\,l^{ab}\,	(M_\pi^2-M_K^2) 
G_2(k^2,M_\eta^2,M_{ss}^2)\,,
\end{eqnarray}
where we assume $M_\eta^2=2B(2m_s+m)/3$
and where
\begin{eqnarray}
G_1(k^2,M^2)&=& 
\frac{1}{(k^2+M^2)}\,,\nonumber\\[-3mm]
\label{eq:finalprops}\\[-3mm]
G_2(k^2,M_1^2,M_2^2)&=&
	\frac {1}{(k^2+M_1^2)(k^2+M_2^2)}\,.\nonumber
\end{eqnarray}
\subsection{Propagators in $SU(4|2)$ chiral perturbation theory}
We consider the case of the effective theory with $N_f=2$ dynamical 
flavours with two degenerate light quarks of mass $m$ and two
 quenched quarks $s$ and $r$ with the mass of the strange quark
$m_s=m_{s_g}=m_r= m_{r_g}$.
The dynamical degrees of freedom in the effective theory have masses
\begin{equation}
M^2_\pi=2 B m\,,\qquad M^2_K=B(m+m_s)\,,\qquad
	M^2_{ss}=2 B m_s\,.
\end{equation}
The propagator $G^{ab}(k^2)$ is defined as follows:
The Lagrangian in Eq.~(\ref{eqn:O2Lag}) restricted to lowest order and with 
$\mathcal{L}_{\rm int}^{(2)}=0$ can be rewritten as
\begin{eqnarray}
\mathcal{L}^{(2)}&=&
\frac 12g^{ab}\left(
	\partial_\mu \phi^a \partial_\mu \phi^b+M_a^2\phi^a\phi^b\right)+
	\frac{1}{6}h^{ab}(M_\pi^2-M_K^2) \phi^a \phi^b\,,
\end{eqnarray}
where
\begin{equation}
M_a^2=\left\{\begin{array}{lll}
M_\pi^2&{\rm for}&a=1,2,3\,,\\
\frac 12\left(M_\pi^2+M_K^2\right)&{\rm for}&a=4,5,6,7,9,10,11,12,16,17,18,19,25,26,27,28\,,\\
M_{ss}^2&{\rm for}&a=8,13,14,15,20,21,22,23,24,29,30,31,32,33,34,35\,,\\
\end{array}\right.
\end{equation}
and where for our choice of the $SU(4|2)$-generators
\begin{equation}
\begin{array}{l@{\hspace{1mm}}c@{\hspace{1mm}}l}
h^{ab}&=&\;\;\;(\sqrt{2}\,g^{a\,8}+ g^{a\, 15}-g^{a\,24}-\sqrt{2}\,g^{a\,35})
	(\sqrt{2}\,g^{b\,8}+ g^{b\, 15}-g^{b\,24}-\sqrt{2}\,g^{b\,35})\,,\\
l^{ab}&=&-(\sqrt{2}\,g^{a\,8}+ g^{a\, 15}+g^{a\,24}+\sqrt{2}\,g^{a\,35})
	(\sqrt{2}\,g^{b\,8}+ g^{b\, 15}+g^{b\,24}+\sqrt{2}\,g^{b\,35})\;.
\end{array}
\end{equation}
In terms of these matrices the propagator is given by
\begin{eqnarray}
\hspace{-1cm}G^{ab}(k^2)&=& 
	g^{ab}G_1(k^2,M_a^2)\nonumber
+	\frac{1}{3}l^{ab} (M_\pi^2-M_K^2)G_2(k^2,M_{ss}^2,M_{ss}^2) \,,
\end{eqnarray}
where $G_1$ and $G_2$ are defined as in Eq.~(\ref{eq:finalprops}).
\section{The integral $\bar B_{21}(q^2,M^2)$}\label{app:B21}
Here we just quote the integral itself. Its derivation from the sum of the
unitary contribution in Eq.~(\ref{eq:unitary}) and the tadpole in 
Eq.~(\ref{eq:tadpole}) can be found for example in the
appendix of Ref.~\cite{Golowich:1995kd}:
\begin{equation}
\bar B_{21}(q^2,M^2)=\frac{1}{12}\left[
	\left(
	1-\frac{4 M^2}{q^2}\right)\bar B(q^2,M^2)-\frac i{48\pi^2}
		\right]\,,
\end{equation}
where for $q^2<4M^2$
\begin{eqnarray}
\bar B(q^2,M^2)&=&-\frac{i}{16\pi^2}\int\limits_0^1
	dx \log\left(
	1-x(1-x)\frac{q^2}{M^2}
	\right)\nonumber\\
	&\stackrel{x=\frac{4M^2}{q^2}}{=}&-\frac{i}{16\pi^2}\left[
	-2+\sqrt{1-x}
	\log\left(
	\frac{\sqrt{1-x}+1}{\sqrt{1-x}-1}
	\right)
	\right]\nonumber\\
	&=&\frac{i}{96\pi^2}\frac{q^2}{M^2}+\frac{i}{960\pi^2}\frac{q^4}{M^4}+
	\dots\,.
\end{eqnarray}
In the body of this paper we found it convenient to absorb a logarithmic term 
into the expression for $\bar B_{21}$:
\begin{equation}
\bar B_{21}(\mu^2,p^2,M^2)=\bar B_{21}(p^2,M^2)-\frac i4 \frac{1}{48\pi^2}
		\log\left(\frac{M^2}{\mu^2}\right)\,.
\end{equation}

\bibliographystyle{JHEP}
\bibliography{spires}

\providecommand{\href}[2]{#2}\begingroup\raggedright\begin{thebibliography}{10}

\bibitem{Neff:2001zr}
H.~Neff, N.~Eicker, T.~Lippert, J.~W. Negele, and K.~Schilling, {\it {On the
  low fermionic eigenmode dominance in QCD on the lattice}},  {\em Phys. Rev.}
  {\bf D64} (2001) 114509, [\href{http://xxx.lanl.gov/abs/hep-lat/0106016}{{\tt
  hep-lat/0106016}}].

\bibitem{Foley:2005ac}
J.~Foley {\em et~al.}, {\it {Practical all-to-all propagators for lattice
  QCD}},  {\em Comput. Phys. Commun.} {\bf 172} (2005) 145--162,
  [\href{http://xxx.lanl.gov/abs/hep-lat/0505023}{{\tt hep-lat/0505023}}].

\bibitem{Gasser:1983yg}
J.~Gasser and H.~Leutwyler, {\it {Chiral Perturbation Theory to One Loop}},
  {\em Ann. Phys.} {\bf 158} (1984) 142.

\bibitem{Gasser:1984gg}
J.~Gasser and H.~Leutwyler, {\it {Chiral Perturbation Theory: Expansions in the
  Mass of the Strange Quark}},  {\em Nucl. Phys.} {\bf B250} (1985) 465.

\bibitem{Bernard:1992mk}
C.~W. Bernard and M.~F.~L. Golterman, {\it Chiral perturbation theory for the
  quenched approximation of {QCD}},  {\em Phys. Rev.} {\bf D46} (1992)
  853--857, [\href{http://xxx.lanl.gov/abs/hep-lat/9204007}{{\tt
  hep-lat/9204007}}].

\bibitem{Bernard:1993sv}
C.~W. Bernard and M.~F. Golterman, {\it {Partially quenched gauge theories and
  an application to staggered fermions}},  {\em Phys.Rev.} {\bf D49} (1994)
  486--494, [\href{http://xxx.lanl.gov/abs/hep-lat/9306005}{{\tt
  hep-lat/9306005}}].

\bibitem{Sharpe:2000bc}
S.~R. Sharpe and N.~Shoresh, {\it {Physical results from unphysical
  simulations}},  {\em Phys. Rev.} {\bf D62} (2000) 094503,
  [\href{http://xxx.lanl.gov/abs/hep-lat/0006017}{{\tt hep-lat/0006017}}].

\bibitem{Jegerlehner:2009ry}
F.~Jegerlehner and A.~Nyffeler, {\it {The Muon g-2}},  {\em Phys. Rept.} {\bf
  477} (2009) 1--110, [\href{http://xxx.lanl.gov/abs/arXiv:0902.3360}{{\tt
  arXiv:0902.3360}}].

\bibitem{Sachrajda:2004mi}
C.~Sachrajda and G.~Villadoro, {\it {Twisted boundary conditions in lattice
  simulations}},  {\em Phys.Lett.} {\bf B609} (2005) 73--85,
  [\href{http://xxx.lanl.gov/abs/hep-lat/0411033}{{\tt hep-lat/0411033}}].

\bibitem{Bedaque:2004ax}
P.~F. Bedaque and J.-W. Chen, {\it {Twisted valence quarks and hadron
  interactions on the lattice}},  {\em Phys.Lett.} {\bf B616} (2005) 208--214,
  [\href{http://xxx.lanl.gov/abs/hep-lat/0412023}{{\tt hep-lat/0412023}}].

\bibitem{Bedaque:2004kc}
P.~F. Bedaque, {\it {Aharonov-Bohm effect and nucleon nucleon phase shifts on
  the lattice}},  {\em Phys.Lett.} {\bf B593} (2004) 82--88,
  [\href{http://xxx.lanl.gov/abs/nucl-th/0402051}{{\tt nucl-th/0402051}}].

\bibitem{deDivitiis:2004kq}
G.~de~Divitiis, R.~Petronzio, and N.~Tantalo, {\it {On the discretization of
  physical momenta in lattice QCD}},  {\em Phys.Lett.} {\bf B595} (2004)
  408--413, [\href{http://xxx.lanl.gov/abs/hep-lat/0405002}{{\tt
  hep-lat/0405002}}].

\bibitem{Tiburzi:2005hg}
B.~C. Tiburzi, {\it {Flavor twisted boundary conditions and the nucleon axial
  current}},  {\em Phys.Lett.} {\bf B617} (2005) 40--48,
  [\href{http://xxx.lanl.gov/abs/hep-lat/0504002}{{\tt hep-lat/0504002}}].

\bibitem{Flynn:2005in}
{\bf UKQCD} Collaboration, J.~Flynn, A.~J{\"u}ttner, and C.~Sachrajda, {\it {A
  Numerical study of partially twisted boundary conditions}},  {\em Phys.Lett.}
  {\bf B632} (2006) 313--318,
  [\href{http://xxx.lanl.gov/abs/hep-lat/0506016}{{\tt hep-lat/0506016}}].

\bibitem{Sharpe:2006pu}
S.~R. Sharpe, {\it {Applications of chiral perturbation theory to lattice
  QCD}},  \href{http://xxx.lanl.gov/abs/[hep-lat/0607016]}{{\tt
  [hep-lat/0607016]}}.

\bibitem{Golterman:2009kw}
M.~Golterman, {\it {Applications of chiral perturbation theory to lattice
  {QCD}}},  \href{http://xxx.lanl.gov/abs/[arXiv:0912.4042]}{{\tt
  [arXiv:0912.4042]}}.

\bibitem{Morel:1987xk}
A.~Morel, {\it {Chiral logarithms in quenched QCD}},  {\em J. Phys. (France)}
  {\bf 48} (1987) 1111--1119.

\bibitem{Blum:2002ii}
T.~Blum, {\it {Lattice calculation of the lowest order hadronic contribution to
  the muon anomalous magnetic moment }},  {\em Phys. Rev. Lett.} {\bf 91}
  (2003) 052001, [\href{http://xxx.lanl.gov/abs/hep-lat/0212018}{{\tt
  hep-lat/0212018}}].

\bibitem{Gockeler:2003cw}
{\bf QCDSF} Collaboration, M.~G{\"o}ckeler {\em et~al.}, {\it {Vacuum
  polarization and hadronic contribution to muon g-2 from lattice QCD}},  {\em
  Nucl.Phys.} {\bf B688} (2004) 135--164,
  [\href{http://xxx.lanl.gov/abs/hep-lat/0312032}{{\tt hep-lat/0312032}}].

\bibitem{Giusti:2008vb}
L.~Giusti and M.~L{\"u}scher, {\it {Chiral symmetry breaking and the
  Banks--Casher relation in lattice {QCD} with Wilson quarks}},  {\em JHEP}
  {\bf 03} (2009) 013, [\href{http://xxx.lanl.gov/abs/arXiv:0812.3638}{{\tt
  arXiv:0812.3638}}].

\bibitem{Juttner:2009yb}
A.~J{\"u}ttner and M.~Della~Morte, {\it {New ideas for g-2 on the lattice}},
  {\em PoS} {\bf LAT2009} (2009) 143,
  [\href{http://xxx.lanl.gov/abs/arXiv:0910.3755}{{\tt arXiv:0910.3755}}].

\bibitem{Weinberg3}
S.~Weinberg, {\em The Quantum Theory of Fields, Volume 3: Supersymmetry}.
\newblock Cambridge University Press, May, 2005.

\bibitem{Kaiser:2000ck}
R.~Kaiser, {\it {Anomalies and WZW-term of two-flavour QCD}},  {\em Phys. Rev.}
  {\bf D63} (2001) 076010, [\href{http://xxx.lanl.gov/abs/hep-ph/0011377}{{\tt
  hep-ph/0011377}}].

\bibitem{Kaiser:2000gs}
R.~Kaiser and H.~Leutwyler, {\it {Large $N_c$ in chiral perturbation theory}},
  {\em Eur. Phys. J.} {\bf C17} (2000) 623--649,
  [\href{http://xxx.lanl.gov/abs/hep-ph/0007101}{{\tt hep-ph/0007101}}].

\bibitem{Bijnens:1999hw}
J.~Bijnens, G.~Colangelo, and G.~Ecker, {\it {Renormalization of chiral
  perturbation theory to order $p^6$}},  {\em Annals Phys.} {\bf 280} (2000)
  100--139, [\href{http://xxx.lanl.gov/abs/hep-ph/9907333}{{\tt
  hep-ph/9907333}}].

\bibitem{Golowich:1995kd}
E.~Golowich and J.~Kambor, {\it {Two loop analysis of vector current
  propagators in chiral perturbation theory}},  {\em Nucl. Phys.} {\bf B447}
  (1995) 373--404, [\href{http://xxx.lanl.gov/abs/hep-ph/9501318}{{\tt
  hep-ph/9501318}}].

\bibitem{Allton:2008pn}
{\bf RBC-UKQCD} Collaboration, C.~Allton {\em et~al.}, {\it {Physical Results
  from 2+1 Flavor Domain Wall QCD and SU(2) Chiral Perturbation Theory}},  {\em
  Phys.Rev.} {\bf D78} (2008) 114509,
  [\href{http://xxx.lanl.gov/abs/arXiv:0804.0473}{{\tt arXiv:0804.0473}}].

\bibitem{Aubin:2006xv}
C.~Aubin and T.~Blum, {\it {Calculating the hadronic vacuum polarization and
  leading hadronic contribution to the muon anomalous magnetic moment with
  improved staggered quarks}},  {\em Phys.Rev.} {\bf D75} (2007) 114502,
  [\href{http://xxx.lanl.gov/abs/hep-lat/0608011}{{\tt hep-lat/0608011}}].

\bibitem{Ecker:1988te}
G.~Ecker, J.~Gasser, A.~Pich, and E.~de~Rafael, {\it {The Role of Resonances in
  Chiral Perturbation Theory}},  {\em Nucl.Phys.} {\bf B321} (1989) 311.

\bibitem{Mainz:g-2}
B.~B. Brandt {\em et~al.}, {\it {Wilson fermions at fine lattice spacings:
  scale setting, pion form factors and $(g-2)_\mu$}},
  \href{http://xxx.lanl.gov/abs/[arXiv:1010.2390] {\rm and M.~Della Morte,
  B.~J{\"a}ger, A.~J{\"u}ttner and H.~Wittig, {\it in preparation}}}{{\tt
  [arXiv:1010.2390] {\rm and M.~Della Morte, B.~J{\"a}ger, A.~J{\"u}ttner and
  H.~Wittig, {\it in preparation}}}}.

\bibitem{Boyle:2007wg}
P.~A. Boyle, J.~M. Flynn, A.~J{\"u}ttner, C.~T. Sachrajda, and J.~M. Zanotti,
  {\it Hadronic form factors in lattice {QCD} at small and vanishing momentum
  transfer},  {\em JHEP} {\bf 05} (2007) 016,
  [\href{http://xxx.lanl.gov/abs/hep-lat/0703005}{{\tt hep-lat/0703005}}].

\bibitem{Jiang:2008te}
F.-J. Jiang and B.~C. Tiburzi, {\it {Flavor Twisted Boundary Conditions in the
  Breit Frame}},  {\em Phys.Rev.} {\bf D78} (2008) 037501,
  [\href{http://xxx.lanl.gov/abs/arXiv:0806.4371}{{\tt arXiv:0806.4371}}].

\bibitem{Renner:2009by}
D.~B. Renner and X.~Feng, {\it {Hadronic contribution to g-2 from twisted mass
  fermions}},  {\em PoS} {\bf LATTICE2008} (2008) 129,
  [\href{http://xxx.lanl.gov/abs/arXiv:0902.2796}{{\tt arXiv:0902.2796}}].

\bibitem{Sharpe:1998xm}
S.~R. Sharpe and J.~Singleton, Robert~L., {\it {Spontaneous flavor and parity
  breaking with Wilson fermions}},  {\em Phys.Rev.} {\bf D58} (1998) 074501,
  [\href{http://xxx.lanl.gov/abs/hep-lat/9804028}{{\tt hep-lat/9804028}}].

\end{thebibliography}\endgroup

\end{document}